\magnification=1200
\hfuzz=3pt
\hsize=12.5cm
\hoffset=0.32cm
\baselineskip=18pt
\voffset=\baselineskip
\nopagenumbers
\font\title=cmssdc10 at 20pt
\font\smalltitle=cmssdc10 at 14pt

\let\text=\textstyle
\let\display=\displaystyle

\def\sectionstyle{\smalltitle}
\newskip\beforesectionskip
\newskip\aftersectionskip
\beforesectionskip=4mm plus 1mm minus 1mm
\aftersectionskip=2mm plus .2mm minus .2mm
\newcount\mysectioncounter
\def\resetsections{\mysectioncounter=0}
\resetsections
\newcount\myeqcounter
\def\mysection#1\par{\par\removelastskip\penalty -250
\vskip\beforesectionskip
\global\advance\mysectioncounter by 1\noindent
\myeqcounter=0{\sectionstyle\the\mysectioncounter.
#1}\par
\nobreak\vskip\aftersectionskip}
\def\myeqno{\global\advance\myeqcounter by 1\eqno{(\the\mysectioncounter.\the\myeqcounter)}}
\def\mydispeqno{\global\advance\myeqcounter by 1\hfill\llap{(\the\mysectioncounter
.\the\myeqcounter)}}
\headline={\hfil\tenrm\folio\hfil}
\footline={\hfil}
\baselineskip=12pt
\parindent= 20pt
\centerline{\smalltitle Aging properties of an anomalously diffusing particule}
\vskip 0.5cm
\centerline{No\"elle POTTIER}
\medskip
\centerline{\sl F\'{e}d\'{e}ration de Recherche CNRS 2438 ``Mati\`{e}re et Syst\`{e}mes Complexes''\/} 
\smallskip
\centerline{\sl and\/} 
\smallskip
\centerline{\sl Groupe de Physique des Solides, CNRS UMR 7588, Universit\'{e}s Paris 6 et Paris 7\/} 

\centerline{\sl 2, place Jussieu, 75251 Paris Cedex 05, France\/}

\vskip 2cm
\noindent
{\smalltitle Abstract}
\medskip
\noindent
We report new results about the two-time dynamics of an anomalously diffusing classical particle, as described by the
generalized Langevin equation with a frequency-dependent noise and the associated friction. The noise is
defined by its spectral density proportional to $\omega^{\delta-1}$ at low frequencies, with $0<\delta<1$ (subdiffusion) or
$1<\delta<2$ (superdiffusion). Using Laplace analysis, we derive analytic expressions in terms of Mittag-Leffler functions for
the correlation functions of the velocity and of the displacement. While the velocity thermalizes at large times (slowly, in
contrast to the standard Brownian motion case $\delta=1$), the displacement never attains equilibrium: it ages. We thus show
that this feature of normal diffusion is shared by a subdiffusive or superdiffusive motion. We provide a closed form analytic
expression for the fluctuation-dissipation ratio characterizing aging.
\vskip 1cm
\parindent=0pt
{\bf PACS numbers:} 

05.40.-a Fluctuation phenomena, random processes, noise and Brownian motion

02.50.Ey Stochastic processes
\vskip 2cm
{\bf KEYWORDS:} 
\medskip
{\parskip=0pt
{\sl Corresponding author\/:
\smallskip
No\"elle POTTIER
  
Groupe de Physique des Solides,

Tour 23, 2 place Jussieu, 75251 Paris Cedex 05, France

Fax number\/: +33 1 43 54 28 78 

E-mail\/: pottier@gps.jussieu.fr\/}}

\vfill
\break
\parskip=0pt
\parindent=20pt
\baselineskip=16pt
\mysection{Introduction}

In this paper, we study the two-time dynamics and the aging properties of a classical particle diffusing in the presence of
coloured ({\sl i.e.\/} frequency-dependent) noise and friction. The dynamics of the particle is governed by the generalized
Langevin equation [1]--[2]. We will not be concerned here with the specific microscopic origin of the noise and of the
friction. The knowledge of the noise spectral density suffices to make the model precise. The colored
noise is conveniently defined by its spectral density, assumed to be proportional to $\omega^{\delta-1}$ at low frequencies
with $0<\delta<1$ or $1<\delta<2$. The frequency-dependent friction is deduced from the noise by means of the second
fluctuation-dissipation theorem (FDT).
 
This general framework can be used in the classical domain as well as in the quantum one, depending on which formulation is
chosen for the FDT. The one-time dynamics of this dissipative model has been extensively studied, with particular emphasis
on the quantum case [3]--[6]. In the classical case, the model allows for the description of anomalous diffusion, that is,
either subdiffusion (for $0<\delta<1$) or superdiffusion (for $1<\delta<2$). Closed analytic expressions of the lowest
moments of the velocity and of the displacement of the anomalously diffusing particle have been obtained in [7]. 

The two-time dynamics of the diffusing particle has been investigated in the case $\delta=1$, which
corresponds to white noise, non-retarded friction and standard Brownian motion [8]--[9]. When studying the two-time correlation
functions of the dynamical variables attached to the particle, one is faced with interesting new physical effects such as
aging, {\sl i.e.\/} the absence of time-translation invariance. Indeed, as
first noticed in [8], the two-time correlation function $\langle x(t)x(t')\rangle$ of the displacement of the diffusing
particle depends explicitly on both times $t$ and $t'$ and not only on the difference $t-t'$, even in the limit of large age
$t'$ ($0\leq t'\leq t$). In contrast to the velocity, which equilibrates at large times, and does not age, the displacement
is an out-of-equilibrium variable. The equilibrium FDT has to be modified in order to relate the displacement response
and correlation functions. This can be achieved through the introduction of a factor (the
fluctuation-dissipation ratio) rescaling the temperature [8]. We have demonstrated in [9] that, for a classical diffusing
particle, the fluctuation-dissipation ratio can be expressed in terms of the time-dependent diffusion coefficients $D(\tau)$
and $D(t_w)$, where $\tau=t-t'$ denotes the observation time and $t_w=t'$ the waiting time. We have subsequently extended this
result to non integer values of $\delta$ between 0 and 2 [10]. 

In the present paper, we study the two-time dynamics of an anomalously diffusing particle. Anomalous diffusion,
which has been observed in various physical systems, is the subject of a growing interest, both experimental and
theoretical (for a recent review see [11]). Making extensive use of Laplace analysis, we derive closed form analytic
expressions for the two-time correlation functions of the velocity and of the displacement. We demonstrate that these two-time
quantities are expressible in terms of Mittag-Leffler functions [12], which have previously been shown to play a central role
in the relaxational one-time dynamics of the particle [7]. As a result, while the particle velocity thermalizes at
large times (slowly, except for $\delta=1$), its displacement never attains equilibrium: it ages. We thus show that this
feature of standard Brownian motion is shared by a subdiffusive or superdiffusive particle. We provide closed form analytic
expressions in terms of Mittag-Leffler functions for the time-dependent diffusion coefficients $D(\tau)$ and $D(t_w)$, as well
as for the fluctuation-dissipation ratio characterizing aging.

\mysection{Generalized Langevin description of anomalous diffusion: a Laplace analysis}

{\bf 2.1. The model and the dynamical variables of interest}

The generalized Langevin equation for a classical particle of mass $m$ diffusing in the absence of a
deterministic potential writes 
$$m\,{dv\over dt}+\int_0^t\gamma(t-t')v(t')\,dt'=F(t),\qquad v={dx\over dt}.\myeqno$$
In Eq. (2.1), $F(t)$ is the Langevin force acting on the particle, as modelized by a stationary
Gaussian random process of zero mean, and $\gamma(t)$ is a retarded friction kernel, which is an even function of $t$ [1]--[2].
As well-known, the coherence of the model implies that
$F(t)$ and
$\gamma(t)$ are not independent quantities, but instead are linked by the second fluctuation-dissipation theorem, that
is
$$\langle F(t)F(t')\rangle=kTm\gamma(|t-t'|),\myeqno$$
where $T$ is the temperature. (The symbol $\langle\ldots\rangle$ denotes the average over the realizations of the noise).
For white noise, one simply has
$$\langle F(t)F(t')\rangle=2kT\eta\,\delta(t-t'),\qquad\gamma(t)=2\gamma\,\delta(t),\myeqno$$
where $\gamma$ is the friction coefficient and $\eta=m\gamma$ the viscosity. The Langevin equation is then non-retarded:
$$m\,{dv\over dt}+m\gamma v=F(t).\myeqno$$

In the following, we shall be interested with the one-time and two-time properties of two dynamical variables attached to the
particle, namely its velocity as defined by the solution of Eq. (2.1) satisfying the initial condition 
$$v(t=0)=v_0,\myeqno$$
and its displacement, as defined by
$$x(t)=\int_0^t v(t')\,dt'.\myeqno$$
We will primarily be concerned with non-white noise of parameters to be specified below and the associated friction
coefficient. Let us beforehand describe the general lines of our analysis, which heavily relies upon the use of Laplace
transformation.
\bigskip
{\bf 2.2. Laplace analysis}

The one-time and the two-time properties of the dynamical variables of interest can conveniently be studied by means of
Laplace analysis, which implies that all the quantities of interest are to be considered as causal functions. Defining, for any
quantity
$q(t)$, its Laplace transform $\hat q(z)=\int_0^\infty q(t)\,e^{-zt}\,dt$,
one gets, by applying the Laplace transformation to the generalized Langevin equation (2.1):
$$mz\hat v(z)+m\hat\gamma(z)\hat v(z)=\hat F(z)+mv(t=0).\myeqno$$
\bigskip
{\bf 2.3. Statistical properties of $\hat F(z)$}

The generalized Langevin model is fully specified by Eq. (2.7) with the initial condition (2.5), together
with the given statistical properties of $\hat F(z)$. Since $F(t)$ is a Gaussian process, the statistical properties of $\hat
F(z)$ are completely characterized by the average value $\langle\hat F(z)\rangle$ and the correlation function $\langle\hat
F(z)\hat F(z')\rangle$.

The Langevin force $F(t)$ being a process of zero mean, one has:
$$\langle\hat F(z)\rangle=0.\myeqno$$
As for the correlation function $\langle\hat F(z)\hat F(z')\rangle$, one gets from the fluctuation-dissipa-tion relation 
(2.2):
$$\langle\hat F(z)\hat F(z')\rangle=kTm\int_0^\infty\!\!\!\int_0^\infty e^{-z't'}\,e^{-zt}\,\gamma(|t-t'|)\,dt\,dt'.\myeqno$$
To compute the double integral (2.9), we shall consider separately the two contributions $\langle\hat F(z)\hat
F(z')\rangle_1$ and $\langle\hat F(z)\hat F(z')\rangle_2$, coming respectively from the time domains $t'\leq t$ and $t\leq
t'$.  Let us first consider the contribution from the domain $t'\leq t$. Setting $u=t-t'$, one gets
$$\langle\hat F(z)\hat F(z')\rangle_1=kTm\int_0^\infty e^{-(z+z')t'}\,dt'\int_0^\infty e^{-zu}\,\gamma(u)\,du,\myeqno$$
that is:
$$\langle\hat F(z)\hat F(z')\rangle_1=kTm\,{1\over z+z'}\,\hat\gamma(z).\myeqno$$
Coming then to the contribution from the domain $t\leq t'$, one gets
$$\langle\hat F(z)\hat F(z')\rangle_2=kTm\int_0^\infty e^{-(z+z')t}\,dt\int_0^\infty e^{-z'v}\,\gamma(v)\,dv,\myeqno$$
that is:
$$\langle\hat F(z)\hat F(z')\rangle_2=kTm\,{1\over z+z'}\,\hat\gamma(z').\myeqno$$
Adding the contributions (2.11) and (2.13), one finally obtains:
$$\langle\hat F(z)\hat F(z')\rangle=kTm\,{\hat\gamma(z)+\hat\gamma(z')\over z+z'}.\myeqno$$
Eq. (2.14), which represents the Laplace domain formulation of the second FDT (Eq. (2.2)), will play a central role in the
following. 

Actually, relations of the form (2.14) hold quite generally. Indeed, given any stationary correlation function of the
form
$$\langle\phi(t)\phi(t')\rangle=A\,f(|t-t'|),\myeqno$$
one can show in a similar way that its double Laplace transform writes
$$\langle\hat\phi(z)\hat\phi(z')\rangle=A\,{\hat f(z)+\hat f(z')\over z+z'}.\myeqno$$
\vfill
\break
{\bf 2.4. Statistical properties of $\hat v(z)$}

One deduces from Eqs. (2.7) and (2.5) the expression of $\hat v(z)$:
$$\hat v(z)={\hat F(z)\over z+\hat\gamma(z)}+{v_0\over z+\hat\gamma(z)}.\myeqno$$
As for the average particle velocity, one has, since $\langle\hat F(z)\rangle=0$:
$$\langle\hat v(z)\rangle={v_0\over z+\hat\gamma(z)}.\myeqno$$
The double Laplace transform $\langle\hat v(z)\hat v(z')\rangle$ of the two-time velocity correlation function $\langle
v(t)v(t')\rangle$ is given by:
$$\langle \hat v(z)\hat v(z')\rangle={\langle\hat F(z)\hat F(z')\rangle\over
m^2[z+\hat\gamma(z)][z'+\hat\gamma(z')]}+{v_0^2\over[z+\hat\gamma(z)][z'+\hat\gamma(z')]}.\myeqno$$
It is the sum of two contributions, the first one coming from the random force correlation function, the second one being
linked to the initial condition. 
Taking into account the expression (2.14) of $\langle\hat F(z)\hat F(z')\rangle$, one can rewrite Eq. (2.19) as
$$\langle\hat v(z)\hat v(z')\rangle={kT\over m}\,{\hat K(z)+\hat K(z')\over z+z'}+
\left(v_0^2-{kT\over m}\right)\,\hat K(z)\,\hat K(z'),\myeqno$$
with $\hat K(z)$ as defined by
$$\hat K(z)={1\over z+\hat\gamma(z)}.\myeqno$$
Formula (2.20) displays clearly the fact that the correlation function $\langle\hat v(z)\hat v(z')\rangle$ is the sum of two
contributions of a different character as far as stationarity properties are concerned. The first one, ${kT\over
m}{\hat K(z)+\hat K(z')\over z+z'}$, of the general form (2.16), corresponds in the time domain to a stationary random
process. The second one,
$\left(v_0^2-{kT\over m}\right)\,\hat K(z)\,\hat K(z')$, corresponds to a function
depending separately on $t$ and $t'$ (and not only on the difference $t-t'$). This term, if non-zero,
will yield an aging contribution to the two-time velocity correlation function $\langle v(t)v(t')\rangle$. 

Averaging out Eq. (2.20) over an equilibrium ensemble of initial velocities, the aging contribution to $\langle
v(t)v(t')\rangle$ disappears. We shall come back to this point later. One is then left with:
$$\langle \hat v(z)\hat v(z')\rangle={kT\over m}\,{\hat K(z)+\hat K(z')\over z+z'}.\myeqno$$
\bigskip
{\bf 2.5. Statistical properties of $\hat x(z)$}

One deduces from Eq. (2.6) the expression of $\hat x(z)$:
$$\hat x(z)={1\over z}\,\hat v(z).\myeqno$$
As for the average particle displacement, one has:
$$\langle\hat x(z)\rangle={1\over z}\,{v_0\over z+\hat\gamma(z)}.\myeqno$$
We assume that the particle velocity is at thermal equilibrium and does not age. The double Laplace transform
$\langle\hat x(z)\hat x(z')\rangle$ of the two-time displacement correlation function
$\langle x(t)x(t')\rangle$ is given by:
$$\langle\hat x(z)\hat x(z')\rangle={kT\over m}\,{1\over z}\,{1\over z'}\,{\hat K(z)+\hat K(z')\over z+z'}.\myeqno$$
Due to the presence of the factors ${1/z}$ and ${1/z'}$ in the r.h.s. of Eq. (2.25), the correlation function $\langle\hat
x(z)\hat x(z')\rangle$ is not of the form (2.16). Consequently, in the time domain, the two-time displacement correlation
function $\langle x(t)x(t')\rangle$ will depend separately on $t$ and $t'$. The particle displacement is thus an aging
variable.

Let us now be more specific and introduce the parameters of the so-called non-Ohmic noise and friction allowing for the
description of the dynamics of a subdiffusive or superdiffusive particle.

\mysection{Noise and friction of the non-Ohmic Langevin model}

We consider a coloured noise of spectral density 
$$\langle|F(\omega)|^2\rangle=
2kT\eta_\delta\,\Bigl({|\omega|\over\tilde\omega}\Bigr)^{\delta-1}\,f_c\Bigl({|\omega|\over\omega_c}\Bigr).\myeqno$$
The small-$|\omega|$ behaviour of $\langle|F(\omega)|^2\rangle$ is a power-law characterized by the exponent $\delta-1$. The
function $f_c({|\omega|/\omega_c})$ is a high-frequency cut-off function of typical width $\omega_c$, and 
$\tilde\omega\ll\omega_c$ denotes a reference frequency allowing for the coupling constant $\eta_\delta=m\gamma_\delta$
to  have the dimension of a viscosity for any $\delta$ [6]. For $\delta=1$, the noise spectral density is a constant
(white-noise), at least in the frequency range $|\omega|\ll\omega_c$. Then the Langevin force is delta-correlated and the
Langevin equation is non-retarded. The white-noise case corresponds to Ohmic friction. The cases
$0<\delta<1$ and $\delta>1$ are known respectively as the sub-Ohmic and super-Ohmic models. Here we will assume that
$0<\delta<2$, for reasons to be developed below [3]--[6]. 

To begin with, we have to compute the correlation function $\langle\hat F(z)\hat F(z')\rangle$. We first deduce from
Eq. (3.1) the expression of the two-time correlation function $\langle F(t)F(t')\rangle$ by using  the Wiener-Khintchine
theorem:
$$\langle F(t)F(t')\rangle=2kT\eta\int_{-\infty}^\infty{d\omega\over
2\pi}\Bigl({|\omega|\over\tilde\omega}\Bigr)^{\delta-1}\,f_c\Bigl({|\omega|\over\omega_c}\Bigr)\,\cos\omega(t-t').\myeqno$$
Then, one deduces from Eq. (3.2) the expression of the double Laplace transform $\langle\hat F(z)\hat F(z')\rangle$:
$$\langle\hat F(z)\hat F(z')\rangle=2kT\eta_\delta\,{1\over
z+z'}\int_0^\infty{d\omega\over
\pi}\Bigl({\omega\over\tilde\omega}\Bigr)^{\delta-1}\,f_c\Bigl({\omega\over\omega_c}\Bigr)\left({z\over
z^2+\omega^2}+{z'\over z'^2+\omega^2}\right).\myeqno$$ This expression is actually of the form (2.14), with
$$\hat\gamma(z)=\gamma_\delta\int_0^\infty{d\omega\over
\pi}\Bigl({\omega\over\tilde\omega}\Bigr)^{\delta-1}\,f_c\Bigl({\omega\over\omega_c}\Bigr)\,{z\over z^2+\omega^2}.\myeqno$$

Since $0<\delta<2$, one can safely make the limit $\omega_c\to\infty$ in formulas (3.3) and (3.4), and write
$$\hat\gamma(z)=\gamma_\delta\int_0^\infty{d\omega\over
\pi}\Bigl({\omega\over\tilde\omega}\Bigr)^{\delta-1}\,{z\over z^2+\omega^2},\myeqno$$
and 
$$\langle\hat F(z)\hat F(z')\rangle=2kT\eta_\delta\,{1\over z+z'}\int_0^\infty{d\omega\over
\pi}\Bigl({\omega\over\tilde\omega}\Bigr)^{\delta-1}\,\left({z\over
z^2+\omega^2}+{z'\over z'^2+\omega^2}\right).\myeqno$$
The integrations over $\omega$ in formulas (3.5) and (3.6) once carried out, one gets
$$\hat\gamma(z)=\gamma_\delta\,\Bigl({z\over\tilde\omega}\Bigr)^{\delta-1}\,{1\over\sin{\delta\pi\over 2}},\myeqno$$
and
$$\langle\hat F(z)\hat F(z')\rangle=kT\eta_\delta\,{1\over z+z'}\left\{\Bigl({z\over\tilde\omega}\Bigr)^{\delta-1}+
\Bigl({z'\over\tilde\omega}\Bigr)^{\delta-1}\right\}\,{1\over\sin{\delta\pi\over 2}}.\myeqno$$

At this stage, following [6], it is convenient to introduce the $\delta$-dependent frequency $\omega_\delta$ as defined by
$$\omega_\delta^{2-\delta}=\gamma_\delta\,{1\over\tilde\omega^{\delta-1}}\,{1\over\sin{\delta\pi\over
2}}.\myeqno$$
Using this notation, one has
$$\hat\gamma(z)=\omega_\delta^{2-\delta}\,z^{\delta-1},\myeqno$$
and
$$\langle\hat F(z)\hat F(z')\rangle=kTm\,\omega_\delta^{2-\delta}\,\left({z^{\delta-1}+z'^{\delta-1}\over
z+z'}\right).\myeqno$$
Eqs. (3.10) and (3.11) together with the Langevin equation (2.7) and the initial condition (2.5) fully specify the model in the
Laplace domain.

\mysection{Particle velocity: one-time and two-time properties}

{\bf 4.1. The average velocity}

By inverting the Laplace transformation, one gets from Eq. (2.18):
$$\langle v(t)\rangle=v_0\,{1\over 2i\pi}\int_{c-i\infty}^{c+i\infty}{1\over z+\hat\gamma(z)}\,e^{zt}\,dz.\myeqno$$ 
The constant $c$ in the definition of the integration contour in Eq. (4.1) is real and chosen such that all the singularities
of $\hat K(z)={1/[z+\hat\gamma(z)]}$ are lying to the left of the integration path. Using the expression (3.10) of
$\hat\gamma(z)$, one is left with
$$\langle v(t)\rangle=v_0\,{1\over 2i\pi}\int_{c-i\infty}^{c+i\infty}{1\over
z+\omega_\delta^{2-\delta}z^{\delta-1}}\,e^{zt}\,dz,\myeqno$$
{\sl i.e.\/}, as showed in [6]--[7]:
$$\langle v(t)\rangle=v_0\,E_{2-\delta}[-(\omega_\delta t)^{2-\delta}].\myeqno$$
Eq. (4.3) displays the fact that the average velocity relaxes towards zero, this decay being described by the
Mittag-Leffler function\footnote{$^1$}{The Mittag-Leffler function is defined by the series expansion
$$E_\alpha(x)=\sum_{n=0}^\infty{x^n\over\Gamma(\alpha n+1)},\qquad\alpha>0,\myeqno$$
where $\Gamma$ is the Euler Gamma function. The Mittag-Leffler function $E_\alpha(x)$ reduces to the exponential $e^x$ when
$\alpha=1$. The asymptotic behaviour at large $x$ of the Mittag-Leffler function $E_\alpha(x)$ is as follows:
$$E_\alpha(x)\simeq-{1\over x}\,{1\over\Gamma(1-\alpha)},\qquad x\gg 1.\myeqno$$}
$E_\alpha(x)$ [12] with $\alpha=2-\delta$ and $x=-(\omega_\delta t)^{2-\delta}$ [6]--[7]. 

At large times ({\sl i.e.\/} $\omega_\delta t\gg 1$), the average particle velocity decreases according to a power-law of time:
$$\langle v(t)\rangle\simeq v_0\,{(\omega_\delta t)^{\delta-2}\over\Gamma(\delta-1)},\qquad
0<\delta<2,\qquad\delta\neq 1.\myeqno$$
In the Ohmic case $\delta=1$, one has $\omega_{\delta=1}=\gamma$. Eq. (4.3) is then nothing but the standard exponential decay
of the Brownian particle average velocity:
$$\langle v(t)\rangle=v_0\,e^{-\gamma t}.\myeqno$$
\bigskip
{\bf 4.2. The velocity correlation function}

Eq. (2.20) yields in the time domain
$$\displaylines{\langle v(t)v(t')\rangle={kT\over
m}\,E_{2-\delta}\bigl[-(\omega_\delta|t-t'|^{2-\delta})\bigr]\hfill\cr
\hfill\hfill +\left(v_0^2-{kT\over m}\right)E_{2-\delta}\bigl[-(\omega_\delta t)^{2-\delta}\bigr]\times
E_{2-\delta}\bigl[-(\omega_\delta t')^{2-\delta}\bigr].\mydispeqno\cr}$$
As expected, the two-time velocity correlation function $\langle v(t)v(t')\rangle$ is the sum of a stationary part and
of an aging one. In the Ohmic case $\delta=1$, Eq. (4.8) reduces to:
$$\langle v(t)v(t')\rangle={kT\over
m}\,e^{-\gamma|t-t'|}+\left(v_0^2-{kT\over m}\right)e^{-\gamma(t+t')}.\myeqno$$

As already noted, the aging part of $\langle v(t)v(t')\rangle$ vanishes if one carries out an average of Eq. (4.8)
over an equilibrium ensemble of initial velocities. Equivalently, this can also be achieved if the initial time, instead of
being taken equal to 0, is taken equal to $t_i$, and subsequently rejected to $-\infty$. The aging part of $\langle
v(t)v(t')\rangle$, which is then equal to the product
$E_{2-\delta}\bigl(-[\omega_\delta (t-t_i)]^{2-\delta}\bigr)\times
E_{2-\delta}\bigl(-[\omega_\delta(t'-t_i)]^{2-\delta}\bigr)$, indeed vanishes in the limit $t_i\to -\infty$. 
In this limit, any initial fluctuation of the average velocity has decayed to zero at finite
time. The particle velocity is then at equilibrium. Correspondingly, the two-time velocity correlation
function reduces to its stationary part:
$$\langle v(t)v(t')\rangle={kT\over
m}\,E_{2-\delta}\bigl[-(\omega_\delta|t-t'|)^{2-\delta})\bigr].\myeqno$$

Interestingly enough, the stationary part of $\langle v(t)v(t')\rangle$ is proportional to the Mittag-Leffler function
$E_{2-\delta}\bigl[-(\omega_\delta|t-t'|)^{2-\delta})\bigr]$ (Eq. (4.10)), while the average velocity
$\langle v(t)\rangle$ resulting from a given initial fluctuation evolves proportionally to $E_{2-\delta}\bigl[-(\omega_\delta
t)^{2-\delta}\bigr]$ (Eq. (4.3)). This fact constitutes the generalization to the non-Ohmic case ($\delta\neq 1$) of the
regression theorem according to which the fluctuations decay in time following the same law as the average value, as it is
well-known for the exponential decay of fluctuations which holds in standard Brownian motion ($\delta=1$).

In the particular case $t=t'$, one gets from Eq. (4.8)
$$\langle v^2(t)\rangle={kT\over
m}+\left(v_0^2-{kT\over m}\right)\,\Bigl\{E_{2-\delta}\bigl[-(\omega_\delta t)^{2-\delta}\bigr]\Bigr\}^2,\myeqno$$
a result which describes the slow relaxation of the second moment of the velocity towards its equilibrium
value [7].

\mysection{Aging of the particle displacement}

From now on, we assume that the two-time velocity correlation function reduces to its stationary part (4.10). The particle
velocity is thermalized. Then, the two-time displacement correlation function can directly be obtained by Laplace inversion of
the correlation function
$\langle\hat x(z)\hat x(z')\rangle$ as given by Eq. (2.25). One thus gets, for $0\leq t'\leq t$:
$$\displaylines{\langle x(t)x(t')\rangle={kT\over m}\,\Bigl\{t^2\,E_{2-\delta,3}\bigl[-(\omega_\delta
t)^{2-\delta}\bigr]+t'^2\, E_{2-\delta,3}\bigl[-(\omega_\delta
t')^{2-\delta}\bigr]\hfill\cr
\hfill -(t-t')^2\,E_{2-\delta,3}\bigl(-[\omega_\delta(t-t')]^{2-\delta}\bigr)\Bigr\},\mydispeqno\cr}$$
where we have used the generalized Mittag-Leffler function\footnote{$^2$}{The generalized
Mittag-Leffler function is defined by the series expansion
$$E_{\alpha,\beta}(x)=\sum_{n=0}^\infty{x^n\over\Gamma(\alpha n+\beta)},\qquad\alpha>0,\qquad\beta>0.\myeqno$$} 
$E_{\alpha,\beta}(x)$ [12]. Another way of deriving the result (5.1) is to compute the double integral
$$\langle x(t)x(t')\rangle=\int_0^t\int_0^{t'}\langle v(t_1)v(t_2)\rangle\,dt_1dt_2\myeqno$$
the correlation function $\langle v(t_1)v(t_2)\rangle$ being taken from Eq. (4.10). One has to use the integration
formulas
$$\int_0^tE_{2-\delta}\bigl[-(\omega_\delta t_1)^{2-\delta}\bigr]\,dt_1=t\,E_{2-\delta,2}\bigl[-(\omega_\delta
t)^{2-\delta}\bigr]\myeqno$$
and
$$\int_0^tdt_1\int_0^{t_1}E_{2-\delta}\bigl[-(\omega_\delta
t_2)^{2-\delta}\bigr]\,dt_2=t^2\,E_{2-\delta,3}\bigl[-(\omega_\delta t)^{2-\delta}\bigr].\myeqno$$
Eq. (5.1) displays the fact that the particle displacement is not a variable at equilibrium. When $\delta=1$, we recover the
result we previously obtained in [9]:
$$\langle x(t)x(t')\rangle={kT\over\eta}\,\left(2t'-{1+e^{-\gamma(t-t')}-e^{-\gamma t}-e^{-\gamma t'}\over\gamma}\right).
\myeqno$$
\bigskip
{\bf 5.1. The equal time correlation function and the time-dependent diffusion coefficient}

In the particular case $t=t'$, one gets from Eq. (5.1):
$$\langle x^2(t)\rangle=2\,{kT\over m}\,t^2\,E_{2-\delta,3}\bigl[-(\omega_\delta t)^{2-\delta}\bigr].\myeqno$$
The time-dependent diffusion coefficient, as defined by [9]
$$D(t)={1\over 2}\,{d\over dt}\langle x^2(t)\rangle,\myeqno$$
is:
$$D(t)={kT\over m}\,t\,E_{2-\delta,2}\bigl[-(\omega_\delta t)^{2-\delta}\bigr].\myeqno$$
In the Ohmic case $\delta=1$, Eqs. (5.7) and (5.9) reduce to the standard formulas
$$\langle x^2(t)\rangle=2\,{kT\over m\gamma}\,t\myeqno$$
and
$$D(t)={kT\over m\gamma}\,(1-e^{-\gamma t}).\myeqno$$

At large times ({\sl i.e.\/} $\omega_\delta t\gg 1$), Eqs. (5.7) and (5.9) yield

$$\langle x^2(t)\rangle\simeq 2\,{kT\over m}\,{1\over\omega_\delta^2}\,{(\omega_\delta t)^\delta\over\Gamma(\delta+1)}\myeqno$$
and
$$D(t)\simeq{kT\over m}\,{1\over\omega_\delta}\,{(\omega_\delta t)^{\delta-1}\over\Gamma(\delta)}.\myeqno$$
Eqs. (5.12) and (5.13) show that, for $\omega_\delta t\gg 1$, the particle motion is subdiffusive for
$0<\delta<1$ and superdiffusive for $1<\delta<2$. In view of this fact, the expression (5.1) of $\langle x(t)x(t')\rangle$
shows that the displacement of a subdiffusive or of a superdiffusive particle is an aging variable, as it is the case for a
normally diffusing particle [8]-[9]. 
\bigskip
{\bf 5.2. The fluctuation-dissipation ratio}

Concerning the particle displacement, which is not a variable at equilibrium, a modified FDT can be written as 
$$\chi_{xx}(t,t')=\beta\Theta(t-t')X(t,t'){\partial\langle x(t)x(t')\rangle\over\partial t'},\myeqno$$
where $\chi_{xx}(t,t')$ is the displacement response function [8]. We have previously demonstrated that, for a
diffusing particle, the fluctuation-dissipation ratio $X(t,t')$ can be expressed in terms of the time-dependent diffusion
coefficients $D(\tau)$ and $D(t_w)$, where $\tau=t-t'$ denotes the observation time and $t_w=t'$ the waiting time [9]-[10]:
$$X(\tau,t_w)={D(\tau)\over D(\tau)+D(t_w)}.\myeqno$$
Using the above found expression (5.9) of the time-dependent diffusion coefficient, one gets:
$$X(\tau,t_w)={\tau\,E_{2-\delta,2}\bigl[-(\omega_\delta\tau)^{2-\delta}\bigr]\over
\tau\,E_{2-\delta,2}\bigl[-(\omega_\delta\tau)^{2-\delta}\bigr]+t_w\,E_{2-\delta,2}\bigl[-(\omega_\delta
t_w)^{2-\delta}\bigr]}.
\myeqno$$
When $\delta=1$, we recover the result we obtained in [9]:
$$X(\tau,t_w)={1-e^{-\gamma\tau}\over 2-e^{-\gamma\tau}-e^{-\gamma t_w}}.\myeqno$$

At large observation and waiting times ({\sl i.e.\/} $\omega_\delta\tau\gg 1$, $\omega_\delta t_w\gg 1$), one can use in Eq.
(5.16) the asymptotic expressions of $D(\tau)$ and $D(t_w)$ as given by Eq.~(5.13). Eq. (5.16) then
displays the fact that, in a subohmic or superohmic model of exponent $\delta$, the large times
aging regime is self-similar, as pictured by the fluctuation-dissipation ratio [10]: 
$$X(\tau,t_w)\simeq{1\over 1+{\Bigl(\display{t_w\over\tau}\Bigr)}^{\delta-1}}.\myeqno$$  
Interestingly enough, $X(\tau,t_w)$ is then a function of ${t_w/\tau}$, solely parametrized by $\delta$.  For
$\delta=1$, one retrieves the result $X={1/2}$ [8]-[9]. For any other value of $\delta$, $X$ 
is an algebraic function of ${t_w/\tau}$. The limits $\tau\to\infty$ and $t_w\to\infty$ do not commute.

\mysection{Conclusion}

We have studied the two-time dynamics of a classical anomalously diffusing particle. The particle motion is
governed by the generalized Langevin equation, with a noise of spectral density proportional to $\omega^{\delta-1}$ at low
frequencies ($0<\delta<1$ or $1<\delta<2$). The Langevin force is modelized by a stationary Gaussian random process. The
velocity and the displacement of the particle are themselves Gaussian processes, fully characterized by their averages and
their two-time correlation functions.

Using a double Laplace transformation technique, we have been able to derive closed form analytic expressions for the
correlation functions of the velocity and of the displacement. Both can be expressed in terms of Mittag-Leffler functions,
which are thus demonstrated to play a central role, not only in the one-time dynamics as shown in [7], but also in the two-time
dynamics of the anomalously diffusing particle. While the velocity thermalizes at large times (slowly, except for the standard
Brownian motion case $\delta=1$), the displacement never attains equilibrium and ages.

The thermal equilibrium velocity fluctuations decay in time following the same Mittag-Leffler law as the average velocity
resulting from a given initial fluctuation. This result constitutes the generalization to the non-Ohmic case $\delta\neq 1$ of
the regression theorem well-known for the exponential decay of fluctuations which holds in standard Brownian motion.

As for the particle displacement, we have obtained for the fluctuation-dissipa-tion ratio characterizing aging
an expression in terms of Mittag-Leffler functions, valid for any values of the observation and waiting times. For
$0<\delta<1$ or $1<\delta<2$, the large times aging regime is described by a self-similar function of ${t_w/\tau}$. This is in
marked contrast to the standard Brownian motion case, in which the large times aging regime is described by the value
$X={1/2}$ of the fluctuation-dissipation ratio.

\vfill
\break
\parindent=0pt
{\smalltitle References}
\bigskip
\baselineskip=12pt
\frenchspacing

1. R. Kubo, {\sl Tokyo Summer Lectures in Theoretical Physics\/} (R. Kubo editor), Benjamin, New York (1966)\kern0.2em;
R. Kubo, Rep. Prog. Phys. {\bf 29}, 255 (1966).

2. R. Kubo, M. Toda and N. Hashitsume, {\sl Statistical physics
\uppercase\expandafter{\romannumeral 2} : nonequilibrium statistical mechanics\/}, Second edition, Springer-Verlag, Berlin 
(1991).

3. P. Schramm and H. Grabert, J. Stat. Phys. {\bf 49}, 767 (1987).

4. H. Grabert, P. Schramm and G.-L. Ingold, Phys. Rep. {\bf 168}, 115 (1988).
 
5. C. Aslangul, N. Pottier and D. Saint-James, J. Physique {\bf 48}, 1871 (1987).

6. U. Weiss, {\sl Quantum dissipative systems\/}, Second edition, World Scientific, Singapore, 1999.

7. E. Lutz, Phys. Rev. E {\bf 64}, 051106 (2001).

8. L.F. Cugliandolo, J. Kurchan and G. Parisi, J. Phys. {\bf 4}, 1641 (1994).

9. N. Pottier and A. Mauger, Physica A {\bf 282}, 77 (2000).

10. A. Mauger and N. Pottier, Phys. Rev. E {\bf 65}, 056107 (2002).

11. R. Metzler and J. Klafter, Phys. Rep. {\bf 339}, 1 (2000).

12. A. Erd\'elyi, {\sl Higher Transcendental Functions\/}, Vol. 3, McGraw-Hill, New-York (1955). 
\bye